\documentstyle[11pt,paspconf,epsf]{article}

\begin{document}

\title{Polars -- multisite emission -- multiwavelength observation}

\author{Axel D. Schwope}
\affil{$^1$ Astrophysical Institute Potsdam, An der Sternwarte 16, D-14482 
Potsdam, Germany}

\author
{K.~Beuermann\altaffilmark{2}, 
D.A.H.~Buckley\altaffilmark{3}, 
D.~Ciardi\altaffilmark{4}, 
M.~Cropper\altaffilmark{5}, 
K.~Horne\altaffilmark{6}, 
S.~Howell\altaffilmark{4}, 
K.-H.~Mantel\altaffilmark{7}, 
A.~Metzner\altaffilmark{1}, 
K.~O'Brien\altaffilmark{6},
R.~Schwarz\altaffilmark{1}, 
M.~Sirk\altaffilmark{8}, 
D.~Steeghs\altaffilmark{6}, 
M.~Still\altaffilmark{6},
H.-C.~Thomas\altaffilmark{9}
}

\affil{}

\altaffiltext{2}{Universit\"{a}ts-Sternwarte G\"ottingen, Geismarlandstr.~11, 
        D-37083 G\"{o}ttingen, Germany}
\altaffiltext{3}{South African Astronomical Observatory, PO Box 9, 
	Observatory 7935, Cape Town, RSA}
\altaffiltext{4}{Department of Physics and Astronomy, University of Wyoming,
	P.O.~Box 3905, University Station, Laramie, WY, 82071, USA}
\altaffiltext{5}{MSSL, University College London, Holmbury St.~Mary, Dorking, 
	Surrey RH5 6NT, UK}
\altaffiltext{6}{Univ.~of St.~Andrews, School of Physics and Astronomy, 
        North Haugh, St.~Andrews, Fife KY16 9SS, Scotland, UK}
\altaffiltext{7}{Universit\"ats-Sternwarte M\"unchen, Scheinerstr.~1, 
        D-81679 M\"unchen, Germany}
\altaffiltext{8}{Center for EUV Astrophysics, University of California, 
	Berkeley CA 94720, USA}
\altaffiltext{9}{MPI f\"{u}r Astrophysik, Karl-Schwarzschild-Str.~1, 85740
	Garching, Germany}

\begin{abstract}
We review the main observational characteristics of AM Herculis stars 
(polars) at X-ray, EUV, UV, IR and optical wavelengths. Particular emphasis
is given to multi-epoch, multi-wavelength observations of the eclipsing 
polar HU Aqr (RX\,J2107.9-0518). 

In AM Herculis stars
the broad-band spectral energy distribution
from X-rays to the IR is governed by only very small structures: the hot
accretion regions on the footpoints af accreting field lines. 
The most extended structures in the binary systems on the other hand, 
the mass-donating secondary stars and the 
accretion streams, distinctly appear only as Doppler-shifted emission or 
absorption lines. They can best be studied by investigating 
selected narrow spectral features in the optical, ultraviolet or the near 
infrared.

In this contribution both aspects will be 
highlighted, the structure of the accretion regions as inferred from 
multi-wavelength observations with low or no spectral resolution, as well
as the structure of the secondary stars and the accretion streams
as inferred from high-resolution spectral observations and Doppler mapping.
\end{abstract}

\keywords{AM Herculis binaries, Doppler tomography, X-rays}

\section{Introduction}
The broad-band spectral energy distribution of polars is governed by the 
processes in the small accretion regions on the footpoints of field lines,
which channels the originally free-falling accretion stream down to the 
white dwarf.
The release of gravitational energy is manifested primarily as bremsstrahlung 
at hard X-rays, quasi-blackbody radiation, 
which is prominent at EUV/soft X-ray 
wavelengths, and as cyclotron radiation, which is detected from the IR to the 
near-UV regime. Details about the relevant processes acting in the accretion 
region and the influence of the main parameters: the specific mass flow 
rate $\dot{m}$ (in g cm$^{-2}$ s$^{-1}$), the magnetic field strength $B$ 
and the 
mass of the white dwarf $M_{\rm wd}$, plus compilations of the 
observational data related to those (X-ray spectra, low-resolution optical 
spectra) have been given in recent reviews by e.g.~Beuermann (1997), 
Beuermann \& Burwitz (1995) and Schwope (1996). In the present paper we 
concentrate on the shape of light curves mainly in the X-ray region.

A different perspective of these systems is possible when viewed through 
ultraviolet ``glasses'', which 
reveal both the heated and unheated parts of the photosphere
of the white dwarf plus reprocessed stream emission. Although observations in 
the ultraviolet have been performed for decades now using IUE, the data 
quality has improved with the advent of HST observation of
polars. We describe here some preliminary results of low spectral resolution 
UV observations, with full phase-coverage of the eclipsing polar HU Aqr.

The existence of accretion streams in polars is a well-established 
observational fact. It derives e.g.~from the broad emission lines in the 
optical/IR/UV with high radial velocities (up to $\sim$2000 km s$^{-1}$) 
varying quasi-sinusoidally, or the absorption dips seen preferentially 
in the X-ray light curves or, more indirectly, from the existence of hot 
plasma in small regions at the white dwarf magnetic poles. However, more direct
information about the size of the streams and the distribution of (luminous)
matter
in the magnetosphere, has only been revealed recently by Doppler 
imaging of a few systems (Diaz \& Steiner 1994, Shafter et al.~1995, Schwope
et al.~1997). Methods which allow us to infer the distribution of luminous 
matter in the magnetosphere are ideal complements to those which allow the 
study of distributions of 'dark' (i.e.~photoabsorbing) matter, which shape the 
EUV/X-ray light curves. We discuss here two examples.

Finally, the donor stars are addressed briefly. We show an example of trailed
spectra of photospheric absorption lines. The Doppler image 
shows a ``half-star'' (i.e.~just one hemisphere) 
only, demonstrating the large effects of X-ray illumination
on the photospheric structure of the secondary star. 

Preliminary results of a broad
multi-wavelength, multi-epoch observational campaign of the eclipsing polar
HU Aqr (RX\,J2107-0518, Hakala et al.~1993, Schwope et al.~1993) provides
the backbone of this paper. Data of similar systems will be shown 
and discussed at appropriate places. A full analysis of the data will appear 
in a series of papers in the near future.

\section{ROSAT and EUVE monitoring of HU Aqr}
\begin{figure}
\begin{minipage}{70mm}
\plotone{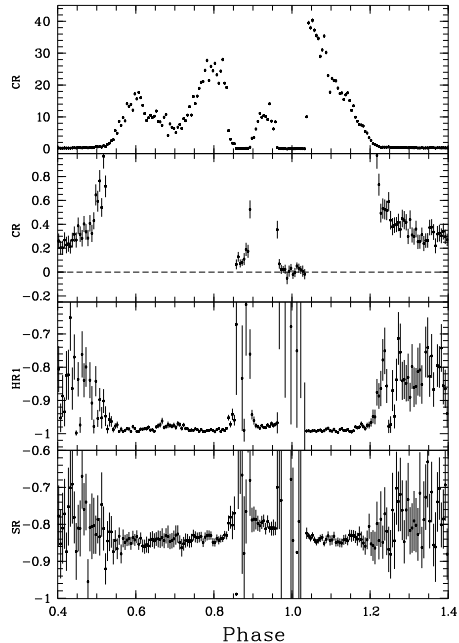}
\end{minipage}
\begin{minipage}{60mm}
\caption{ROSAT PSPC X-ray light curve of HU Aqr obtained in a 
high accretion state in October/November 1993. Shown are from top to bottom 
the phase-averaged light curve in the total ROSAT band (enlarged scale in 
second panel), the hardness ratio and softness ratio as defined in the text}
\label{hu_pspc}
\end{minipage}
\end{figure}

In Fig.~\ref{hu_pspc} we show results of a 37 ksec observation of HU Aqr 
with the PSPC onboard ROSAT, when the system was in a high accretion state.
ROSAT X-ray spectra of polars usually show two components: a soft component, 
$S$, below 0.5 keV and a hard bremsstrahlung component, $H$, 
above 0.5 keV. The hardness ratio HR1, defined as $(H-S)/(H+S)$, balances both 
components with respect to each other. The softness ratio SR is defined in an
equivalent way by defining two subchannels $S1$ and $S2$ in the soft band $S$. 
HR1 and SR are also shown in Fig.~\ref{hu_pspc}.

%\begin{figure}[t]
%\plotone{pspc_cr_hr_sr_ps}
%\plotfiddle{pspc_cr_hr_sr_ps}{9cm}{0}{40}{40}{-120}{-20}
%\label{hu_pspc}
%\end{figure}

\begin{figure}[t]
\plottwo{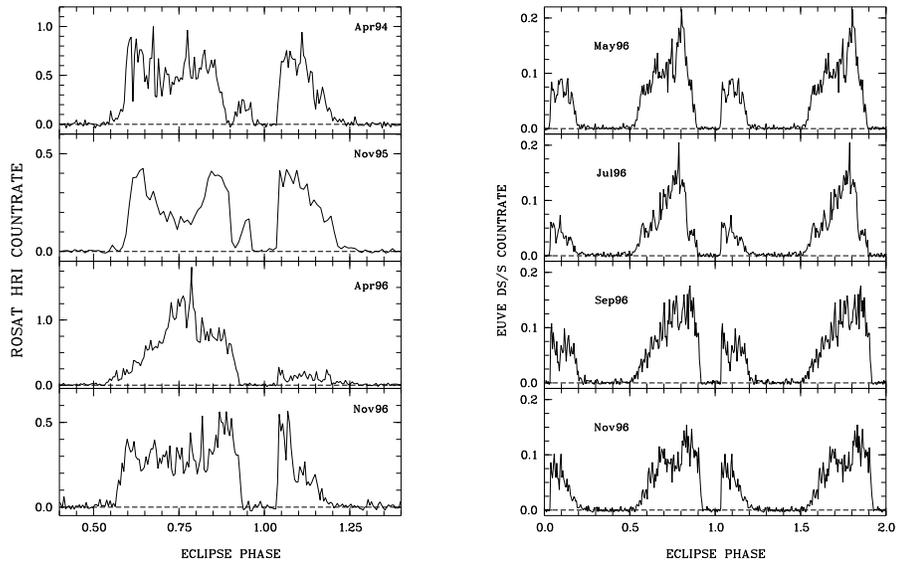}{euve_4epochs_lcs_ps}
\caption{X-ray and EUVE light curves of HU Aqr obtained in states of reduced 
accretion (epochs of observation are labelled in the individual panels). 
Typical observation time with the HRI was 20 ksec, with EUVE 90 ksec. All 
data where phase-folded and averaged using the preliminary ephemeris
given by Schwope et al.~(1997).} 
\label{hu_hri_euve}
\end{figure}

The X-ray light curve of Fig.~\ref{hu_pspc} shows several distinct 
features which will be discussed in the following paragraphs. We mention:
(1) the existence of bright and faint phases alternating with the orbital 
period due to the specific location of the one main accretion region on the 
synchronously rotating white dwarf, (2) an eclipse by the secondary star
centered on phase $\phi = 0.0$, (3) a pre-eclipse dip between $\phi = 0.85$
and 0.90 caused by the intervening magnetically trapped 
part of the accretion stream,
(4) accretion flares during the X-ray bright phase caused by the impact of 
overdense (denser than the average) clumps of matter (not so obvious from the 
phase-averaged representation of Fig.~\ref{hu_pspc}, but better visible in some
panels of Fig.~\ref{hu_hri_euve}a, the phase-averaged ROSAT HRI light curves),
(5) a broad flux depression between phase $\phi = 0.60$ and 0.77 and (6) 
the clear
detection of X-rays during the nominal faint phase, when the accretion spot 
is out of view. In addition, Sohl et al.~(1995) report the 3$\sigma$ detection 
of HU Aqr during eclipse phase (using the same data), which they assign to
the corona of the secondary star, the first such detection in an AM Herculis 
star.

Faint phase X-ray emission was reported also for other self-eclipsing 
systems as e.g.~VV Pup (Schwope et al.~1995a) and QS Tel (Schwope et 
al.~1995b). Since both these systems are well-known
two-pole accretors, from the detection of cyclotron lines from both poles, 
faint-phase X-rays were assigned to these secondary poles. In these
two systems, as well as in HU Aqr, the faint-phase X-ray emission is 
significantly harder than the softer bright-phase emission. HU Aqr has not
shown signs of a second active pole, like circular polarization or 
cyclotron humps. Unless the magnetic field strength is in some sense 
unusual (high or low), which would make cyclotron radiation non-detectable
in the optical 
or the ultraviolet, a different explanation has to be found for the faint-phase
X-ray emission in HU Aqr (which then may also be relevant for the two stars 
mentioned). Obvious candidates are scattered X-rays from either the 
stream or the secondary star. Since stream emission should be detectable 
also during the first few minutes of the eclipse (unless these X-rays are 
blocked by the accretion curtain, see below), scattered X-rays from the 
secondary seems to be the more likely choice.

Broad flux depressions as that of Fig.~\ref{hu_pspc} ($\phi = 0.60 - 0.77$)
are seen in several other systems, most pronounced 
in the EUVE light curve of UZ For (Warren et al.~1995), in several respects 
a twin to the PSPC light curve of HU Aqr. The hardness ratio HR1 increases 
slightly at that phase, indicating that absorption takes place. This helps
to understand these depressions as caused by absorption (instead of missing
emission). The huge width of the dip requires the 
absorbing region to be located in the very vicinity of the emission source.
It's explanation requires a certain temperature and geometric structure
of the accretion region as well as a specific viewing geometry. Sirk \& 
Howell (1997) 
have developed a numerical model and applied it to the EUVE light curves
of several polars. However, such structures are obviously 
non-stationary. A look on the ROSAT HRI light curves of HU Aqr obtained at 
several epochs during recent years (Fig.~\ref{hu_hri_euve}a) shows the broad 
dip to be clearly present only again in Nov95, whereas a flux maximum is 
observed at the same phase half a year later (Apr96). At two other occasions 
(Apr94, Nov96) the light curve is roughly box-shaped and dominated by flares. 
These latter examples  are quite reasonably reflected by the Hameury \& King
(1988) `blobby accretion model' with no additional absorption. This model 
was originally developed in order to explain the anomalous state of 
AM Her itself. 

Of particular interest is the variability of the pre-eclipse dip. Since
HU Aqr is a high inclination system, $i \simeq 85\deg$, its phase and width
are good tracers of the location and extent of the coupling region, where
the initially free-falling stream is redirected. The different X-ray 
observations show clearly that this region is continously moving towards 
$L_1$ (the dip approaches the eclipse). During 1993--1995 the dip and the 
eclipse
appeared as separate structures, while 
at later epochs both features are merged.
Even at the early dates the X-ray flux between dip and eclipse was reduced 
to about 1/3 of the post-eclipse flux, which indicates the presence of an 
accretion curtain built by matter tugged from the whole ballistic stream 
between $L_1$ and the coupling region, where finally 
the bulk of matter is threaded.
The EUVE light curves (Fig.~\ref{hu_hri_euve}b) show the same trend of 
dip motion on a time scale as short as two months. 

\begin{figure}
\begin{minipage}{2.in}
%\plotone{stagnation_ps}
\plotfiddle{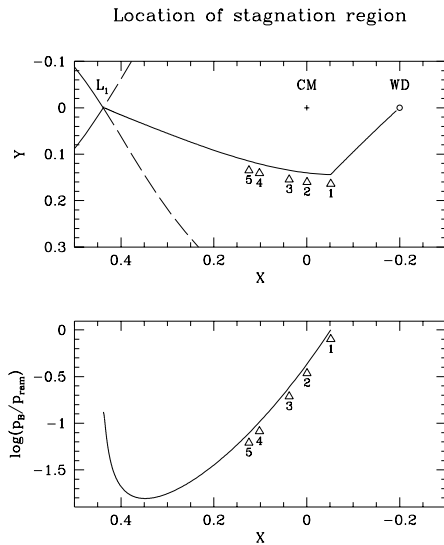}{2.in}{0}{40}{40}{-88}{-98}
\end{minipage}
\begin{minipage}{3.25in}
\caption{Model computations for the accretion stream in HU Aqr. In the upper 
panel a sketch of the stream and binary geometry is shown. The mass-donating
companion star is to the left, the accretion stream starts at 
$L_1$. Initially it follows a ballistic trajectory, later
it is re-directed by the magnetic field. Triangles mark the location of the
stagnation or coupling region as measured from dip-phases at different 
occasions (1 -- Oct 1993 \dots\ 5 -- Nov96). 
In the lower panel the ratio between magnetic and ram pressure along the
ballistic stream is shown. The computations assume constant cross section
$\rho v = \dot{m}=const$ along the stream.}
\label{hu_coup}
\end{minipage}
\end{figure}

We have modelled the stream with a single-particle code. It is assumed that 
the stream follows a free-fall trajectory as long as 
the ram pressure $p_{\rm ram} \propto \rho v^2 = v \dot{m}$ 
of the plasma is larger than the magnetic pressure $p_{B} \propto B^2$.
The latter is computed assuming a dipolar field geometry with
polar field strength $B = 37$\, MG and appropriate orientation in order to 
reflect the location of the accretion spot on the white dwarf in the high 
accretion state (azimuth derived from the center of the bright phase, 
co-latitude derived from the motion of cyclotron harmonics). Results of this
calculation are shown in Fig.~\ref{hu_coup}. 
This simple model gives surprisingly 
good agreement with the observations for the high state (epoch '1'). The 
magnetic pressure equals the ram pressure at an azimuth of about $45\deg$,
as observed. A pronounced decrease of the ram pressure by a factor of 
$\sim$10 is required in the framework of this model in order to shift the
coupling point from '1' to '5'. The reduced velocity at '5' may account 
for a factor $\sim$2, the remaining factor 5 requires an explanation in 
terms of reduction of $\dot{m}$
(hence, of $\dot{M}$, the total mass flow rate in g s$^{-1}$). 
Indeed, a large decrease of the X-ray flux has taken place between 
epoch '1' (PSPC) and epochs '2--5' (HRI).
The large reduction in count rate can be accounted for only partially 
by the reduced sensitivity of the HRI with respect to the PSPC, the conversion
factor for a soft source as HU Aqr is about 5.5, hence there is a real 
decrease of X-ray flux by a factor of $\sim$5. Interestingly, 
there is no obvious continous further 
decrease of the HRI countrate between epochs '2' 
and '5'. This means that either our simple picture of stream-field interaction
is incomplete or that the HRI-countrate cannot be used as unique tracer of
$\dot{M}$, because of its lack of 
energy resolution. Such data may strongly be affected by absorption, or 
alternatively a significant part of the X-ray flux might be shifted out of the
observation window due to temperature changes.

Although the azimuth of the coupling region is shifted by more than $20\deg$,
no corresponding shift of $> 0.05$ phase units of the bright phase center
with respect to eclipse center is observed. This might be a hint to a 
complicated field structure in the vicinity of the white dwarf surface, where
higher multipoles might become important.

\section{Accretion mode changes in BL Hyi}
\begin{figure}[t]
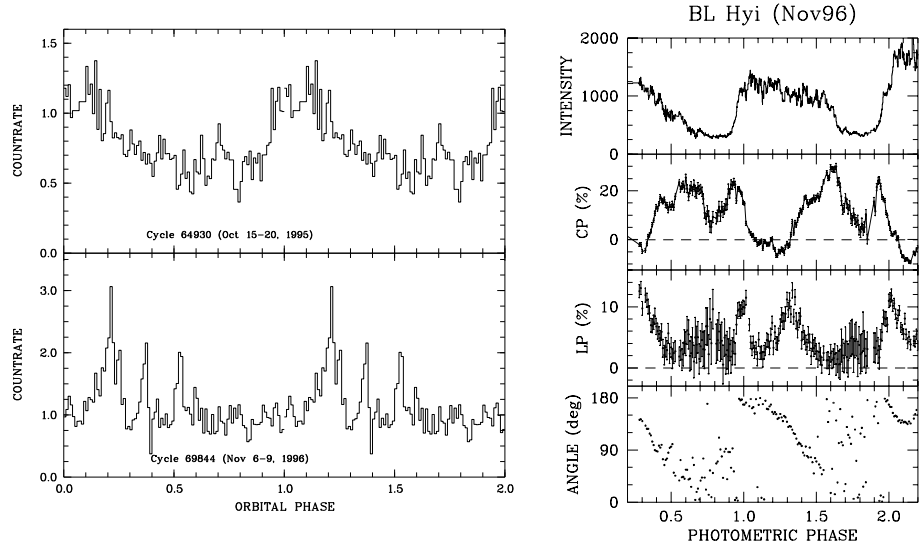

\begin{minipage}{2.6in}
\plotfiddle{bl_hri_ps}{6cm}{0}{40}{40}{-110}{-10}
\end{minipage}
\begin{minipage}{2.6in}
\plotfiddle{bl_pol_ps}{6cm}{0}{50}{50}{-100}{-90}
\end{minipage}
\caption{ROSAT X-ray (left) and optical photometric/polarimetric light
curves (right) of BL Hyi obtained during high accretion states in 1995 
and 1996. The 1996 data were taken quasi-simultaneously.}
\label{bl_hri_pol}
\end{figure}

Another striking case for large variations in the accretion geometry is
BL Hyi. It was monitored already in the EXOSAT era by Beuermann \& Schwope
(1989) and at that time mostly showed a simple, i.e.~single humped, 
light curve,
suggesting that only one pole (the `southern' pole,
below the orbital plane) was 
active. However, occasional soft flares in the nominal faint phase 
were suggestive of a second active region. In the 1980's and during the 
early 1990's BL Hyi was accreting for most times at low rates. In the 
meantime (since $\sim$1993) the system has returned to its active state 
similar to that at discovery some 15 years ago. These changes have dramatically
influenced  the shape of the X-ray light curve, for which 
we show the two most recent
examples in Fig.~\ref{bl_hri_pol}, together with phase-resolved photoelectric
polarimetry obtained quasi-simultaneously in 1996. A hump indicating the
start of the nominal bright phase can still be recognized in 1995. However,
it has not survived the ongoing changes in the accretion geometry, since the 
1996 phase-folded light curve is almost flat, with strong flares superimposed.
This is contrary to what is seen in the optical, where the photometry
still reveals a pronounced hump.
Only the polarimetry reveals clearly that a second pole is active 
by the detection of highly polarized 
light in the nominal faint phase with opposite sign to the bright phase.

The accretion geometry and the field topology of BL Hyi 
are still a mystery. Cropper (1987) 
and Piirola et al.~(1987) favour a high inclination as large as $75\deg$, 
wheras Schwope \& Beuermann (1989) argue for a low inclination of 
$\sim$$35\deg$. Ferrario at al.~(1996) find a field strength of 23\,MG in the
accretion plasma, Schwope et al.~(1995) deduce a low value of 12\,MG and 
propose a field configuration with strong multipole components. In summary, 
BL Hyi 
is after all these years a puzzling system and observers need to sharpen their
tools in order to uncover the basic parameters of this sytem, for example one 
may think of Doppler imaging of the accretion stream and/or the secondary star.

\begin{figure}[t]
\plotfiddle{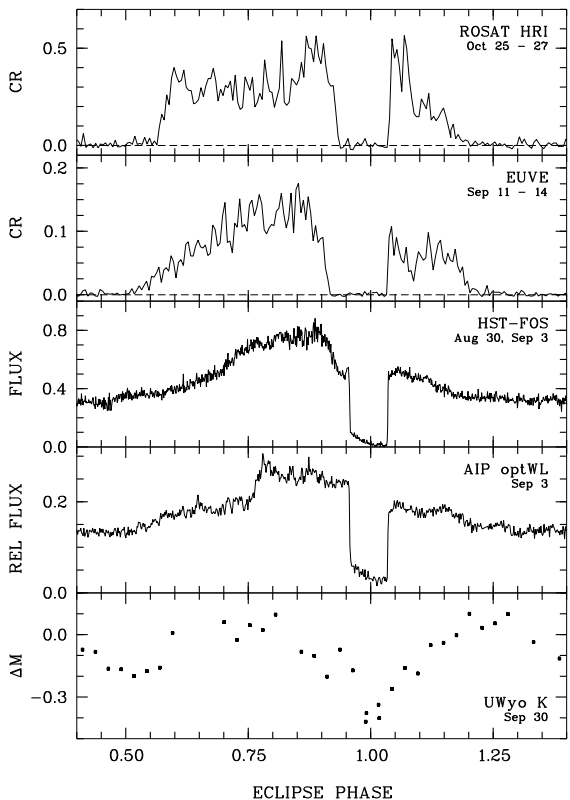}{9.5cm}{0}{45}{45}{-135}{-40}
\caption{Phase-folded light curves of HU Aqr obtained in different
wavelength bands during a coordinated campaign in autumn 1996. 
The instruments and/or wavelength bands are indicated in the 
individual panels.}
\label{hu_mult}
\end{figure}

\section{Multiwavelength photometry of HU Aqr}
The large changes of accretion modes which are observed in several 
well-studied polars underline the importance of near-simultaneous 
multiwavelength observations. An example is the  data collected 
on HU Aqr in Sep--Oct 1996, which is shown in Fig.~\ref{hu_mult}.
These observations were obtained from space with
ROSAT, EUVE and HST and from ground with the AIP 70cm telescope (white light
optical photometry) and with the WIRO (Univ.~Wyoming 2.3m IR telescope).

The structure of ROSAT and EUVE light curves of HU Aqr have been discussed 
above. Here we emphasize one further point, the size of the accretion 
curtain. The HRI onboard ROSAT and the EUVE DS/S are sensitive mainly to the
same radiation component: the soft quasi-blackbody component from the 
accretion spot. Both probe this component at somewhat different
energies, with the EUV range being much more sensitive to photo-absorbing 
cold matter. This is the reason that the dip ingress occurs much earlier
in phase in the EUV than in the soft X-ray range. Although the individual
instruments do not possess energy resolving power, their combined scans
of the dip ingress help to constrain the density profile of the 
accretion curtain. Appropriate modelling is in progress.

In the third panel of Fig.~\ref{hu_mult} the mean flux in the 1200-2500\,\AA\
range is shown as seen by the HST-FOS. The ultraviolet is dominated by 
emission from the white dwarf (at all phases outside eclipse), the heated 
photosphere in and around the accretion region (bright phase), and the 
accretion stream. The latter is visible at all phases, but in particular 
during the first half of the eclipse. An absorption dip
is present in the UV too, it is not seen in the optical, contrary to the 
high state where a pronounced dip was seen in UBVRI light curves 
centered at phase 0.88
(Schwope et al.~1997). Temperatures of the white dwarf and the warm UV-spot
were tentatively determined from the faint-phase and bright-minus-faint-phase
spectra, and found to be $T_{\rm wd} \sim 10000$\,K and $T_{\rm spot}
\sim 35000$\,K. The smooth rise and fall of the UV flux to and from maximum
flux can be modelled with a flat spot on the surface on the white dwarf
(approximate location at co-latitude $27\deg$). The 
light curves at higher energies, which are thought to originate from small 
regions embedded in the UV-spot, show a much steeper rise and fall which is
not compatible with a flat spot and leads to the conclusion that some vertical 
extent of the emission region is necessary.

The bright phase in the optical is dominated
by cyclotron radiation from a $\sim$10\,keV plasma.
Again, contrary to the high state, the optical light curve is single-humped 
only (see Schwope 1995 for a comparison of high- and low-state light curves). 
This suggests a much more structured 
emission region present in the low state than the relatively simple 
(in terms of cyclotron beaming)
and repeatable variation of optical flux suggested for the high state.

\begin{figure}[t]
\plottwo{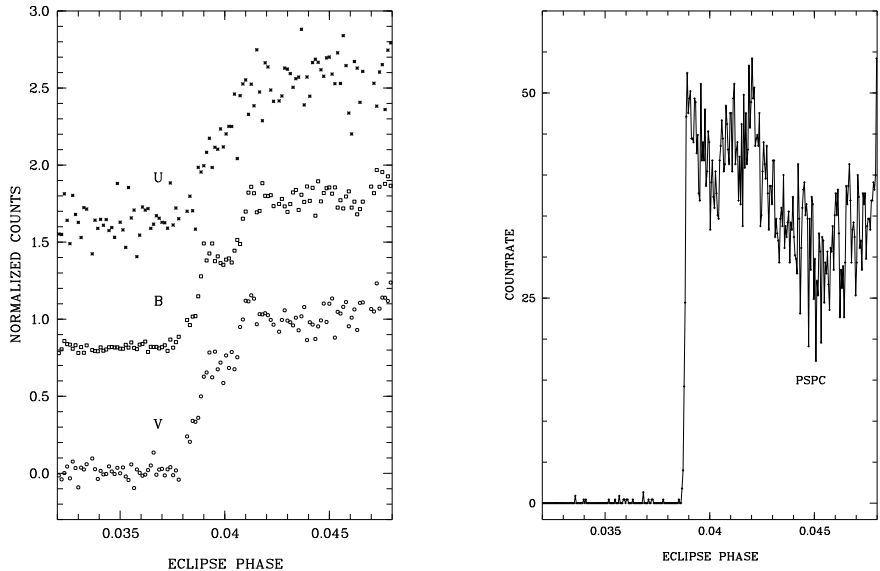}{pspc_egr_large_ps}
\caption{Phase-folded eclipse light curves of HU Aqr centered on egress
phase obtained in August 1993 (optical UBV, left) and October/November 1993
(ROSAT, right). The X-ray observations were
affected by the wire meshes of the PSPC detector, hence, possible substructure
might be instrumental.}
\label{hu_ecl}
\end{figure}

Finally, the IR reveals the secondary star by its ellipsoidal light 
curve. This causes two humps separated by 0.5 in phase. However, the eclipse 
is still present at IR wavelengths which indicates the IR-tail of the 
cyclotron component from the accretion spot (for details of the IR-data see
the Ciardi et al.~(1997)).

Information about the sizes and other parameters of the different emission 
components can be extracted from  eclipse light curves.
We show in Fig.~\ref{hu_ecl} two data sets obtained 2.5 months apart 
with high time resolution in the optical and in the X-ray range.
The phase interval covered by Fig.~\ref{hu_ecl} corresponds to
two minutes and covers the egress of the white dwarf (plus hot spot) 
completely. The egress of the accretion stream is complete only after 
phase 0.1, hence, it is not included in Fig.~\ref{hu_ecl}. 
The data shown here for HU Aqr can be compared with those
of its twin, UZ For (see e.g.~Bailey \& Cropper 1991 for optical data, 
Stockman \& Schmidt 1996 for UV data and Watson 1993 for optical/X-ray data).
As in UZ For, the X-ray spot is small, egress takes only about 1\,sec.
The X-ray egress seems to be essentially simultaneous with the fastest part
of the optical egress. A deconvolution of the optical light curve is 
difficult, at least four components have to be taken into account:
the white dwarf, the spot on it (most prominent in the UV), cyclotron radiation
and the accretion stream. Particularly 
intriguing is the standstill (or even decrease) of optical flux between 
$\phi =0.039$ and 0.0405, which is prominent in $B$ and $V$, but not in $U$. 

\begin{figure}[t]
\plotfiddle{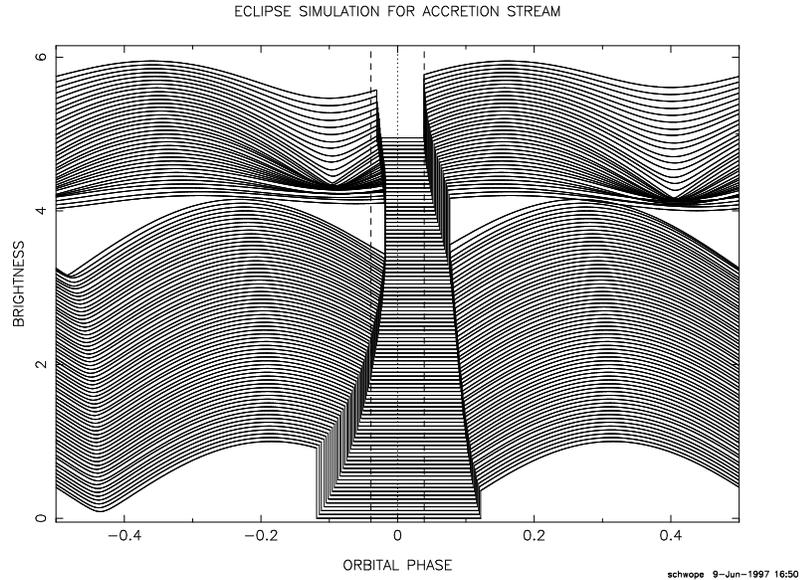}{54mm}{-90}{40}{40}{-150}{230}
%\plotfiddle{stream_lc_high_thick_ps}{54mm}{-90}{40}{40}{-150}{230}
%\plotone{stream_el_high_thick_ps}
\caption{Simulated light curves of stream elements
along the one-particle trajectory shown in Fig.~3 (upper panel) in the 
optically thick approximation. Vertical dashed lines indicate the
observed eclipse width and eclipse center of HU Aqr. 
Constant arbitrary temperature has been assumed along the stream. The light 
curves of individual elements have been plotted with a small vertical offset
with respect to each other, starting with the element nearest to $L_1$ at 
bottom. The jump in intensity at element $\sim$55 occurs when the stream 
becomes aligned with the magnetic field (coupling region) and the projection 
angle changes over a short distance.
}
\label{ecl_sim}
\end{figure}

The brightness distribution along the accretion stream in HU Aqr 
has been mapped by Hakala (1995) using low-state eclipse data and applying 
a MEM-code combined with genetic optimisation. We have started trying a 
supplementary approach by synthesizing light curves (and trailed spectrograms)
for an assumed geometry of the stream. The stream is divided in typically 
100 elements, optically thick emission is assumed and,
for the first generation models,
constant temperature along the stream. Individual elements contribute 
to the total signal depending on the foreshortening angle and visibility.
An example of such a calculation is shown in Fig.~\ref{ecl_sim}. 
Stream parameters were
chosen in order to match the likely situation in HU Aqr in its high state.
This example shows, that the length of eclipse ingress phase is determined
by the extent of the horizontal (ballistic) stream, not by the vertical 
(magnetic) stream. The situation in the low state is presently unsolved. The 
sum of the contributions of all stream elements shown in Fig.~\ref{ecl_sim} 
results in a double-humped light curve with equal maximum height occuring
 at phases $\sim$0.25 and $\sim$0.75, similarly to what is observed in the 
1993 high state (Schwope et al.~1997 show high-resolution light curves 
of the main emission lines). The observations show in addition, that the stream
is brighter on its illuminated concave side. The amount of this brightness 
contrast depends on the line concerned.  It is higher for the optically 
thicker Balmer lines, lower in the (still optically thick) line of ionized
Helium He{\sc ii}\,$\lambda 4686$.

\begin{figure}[t]
\plotfiddle{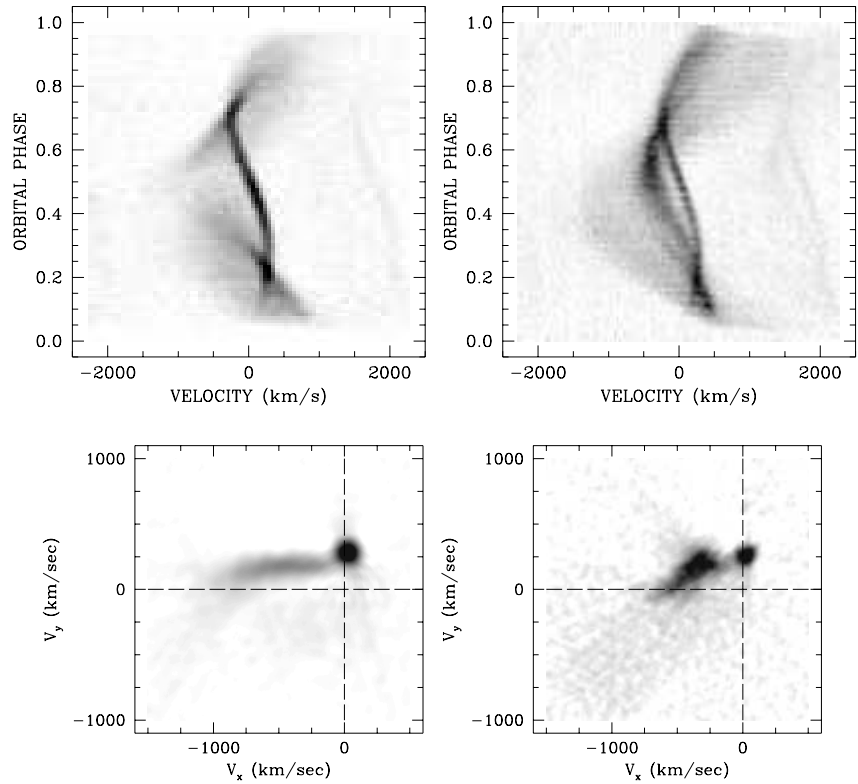}{10cm}{0}{85}{85}{-260}{-270}
\caption{Trailed spectrograms of the HeII 4686 emission line of HU Aqr 
obtained in a high state (1993, left) and a state of reduced accretion (1996, 
right). The trailed spectra are continuum-subtracted.
In the lower panels the Doppler maps of the trailed spectrograms are shown
computed by filtered backprojection.
}
\label{hu_trails}
\end{figure}

\section{Doppler imaging of HU Aqr - uncovering the stream}
In  recent years the availability of low-noise CCDs with sufficient high 
quantum efficiency
allowing high-time resolution spectroscopy of even short-period CVs (phase
resolution better than 0.05 required) opened a new window on these systems
through 
detailed studies of their emission lines. When analysed by Doppler mapping
(Marsh \& Horne 1988), images of the line emission regions
in the binary system in velocity space $(v_x, v_y)$ can be constructed. A first
such investigation of a polar, VV~Pup, was presented by Diaz \& Steiner (1994).

We obtained high-resolution spectra with full-phase coverage of HU Aqr
at two epochs. The first data set was taken in the 1993 high state, the 
second in a state of reduced accretion in 1996 (Fig.~\ref{hu_trails}). The 
effect of the reduced accretion rate on the trailed spectrograms is quite
dramatic. A constant feature at both occasions is the sinusoidally varying 
narrow emission line component (NEL) 
from the heated face of the companion star.
This component is bright in the trails at $\phi =0.5$, when the frontside of 
the secondary is best visible. The Doppler image of this component is the 
bright spot at $(v_x, v_y) \simeq (0, 290)$ 
km s$^{-1}$. 
These spots, however
do not appear exactly at $v_x = 0$ km s$^{-1}$, as they should if illumination 
and re-radiation would occur homogeneously. The shift towards positive 
velocities in $v_x$ (towards the trailing edge of the secondary star, right in 
the Doppler map), indicates that significant shielding of the leading 
hemisphere takes place (Schwope et al.~1997). This is another observational 
clue to the presence of an accretion curtain, apart from the reduced
flux between the narrow dip and the eclipse in the X-ray light curves.

Two further emission line components are visible in the high-state 
trailed spectrum, another rather narrow so-called high-velocity component (HVC)
crossing the NEL at phases $\phi = 0.22$ and 0.70, and a broad 
underlying component (BBC). These can be assigned to the two parts of the 
accretion stream, the ballistic (horizontal) part and the magnetic (vertical)
part. The correspondence, however, is not exact.
The HVC is not as bright as the BBC in the trailed spectrogram, but in the map 
the horizontal stream (upper left quadrant) is much brighter than the vertical
stream (lower left quadrant). This results from the somewhat artificial 
division between HVC and BBC.

Emission from the horizontal stream is expected to be spread over a large
wavelength interval at phases $\sim$0.05 and $\sim$0.55, when the stream
is most directly moving away from or approaching the observer. The observer
then sees the differently accelerated parts of the stream. If the stream
is significantly optically thick, intensity minima should occur also at 
these phases. Vice versa, around quadrature phases the stream passes the 
observer with essentially no significant radial velocity (apart from the 
orbital velocity). Consequently, the velocity spread is small, a narrow
emission features, the HVC, emerges. 
If emission is optically thick, this emission
line component is also bright, since the observer sees the stream in 
full elongation with smallest foreshortening.

All these features can clearly be recognized in the trailed spectrogram
obtained in the 1993 high state. The Doppler image shows the horizontal
stream to be extended down to $(v_x, v_y) \simeq (-1000, 0)$ km s$^{-1}$. 
Such high velocities are reached if the coupling region lies sufficiently 
downstream (marked by '1' in Fig.~\ref{hu_coup}). The X-ray light curves
indicate that at reduced accretion rate the coupling region is shifted 
towards the inner Lagrangian point $L_1$. Is this evident from the trailed
spectrograms and the Doppler images, too? 

On the first glance this is perhaps not so evident from the trailed 
spectrogram alone but the Doppler map shows clearly that the horizontal 
stream is far from being so elongated as it has been in the high state. 
It extends to only about $v_x \simeq -500$ km s$^{-1}$.
Relative to the high state, the vertical stream (lower left quadrant)
is more intensive, which is (at least partially) due to the fact, that
it is more extended.

The trailed spectrogram shows that both the components originating from the
stream  changed considerably. The BBC is much more easily recognizable at high 
velocities than in the high state and the HVC lies much closer to the NEL 
than in the high state. The latter is a direct consequence from the 
horizontal stream being much less elongated. There is still a clear separation 
between NEL and HVC between phases 0.3 and 0.6, indicating that the 
stream at $L_1$ is fainter than further downstream. The HVC shows some 
indication for splitting between phases 0.2 -- 0.4. The ultimate reason 
for this behaviour is presently unclear, it might indicate the presence 
of a second accretion stream. 

High-resolution high-speed spectroscopy in addition opens a new dimension 
for studies of the distribution of matter in the magnetosphere since it
allows to establish eclipse light curves in velocity bins (and eclipse
mapping in velocity bins). Such studies have so far never been tried 
in polars, data as those presented here might open the door to the field.

\section{Doppler imaging of QQ Vul - uncovering half stars}
The trailed spectrograms of HU Aqr (and those of most AM Herculis stars)
clearly display a narrow emission line component which is attributed to the
secondary star. A lot of valuable information about the binary system and the 
stellar components can be extracted from observations of the NEL using its 
repeatability, its radial velocity, its width (including variability), and 
intensity (including variability) to derive, e.g., the orbital period, the 
stellar masses, the size of the secondary star, and the power of the 
illuminating X-ray source. All but the first point mentioned require a 
good estimate of the effects of illumination on the secondary. We have 
already seen, that illumination/re-radiation is (or might be) far from what
a numerical model would usually assume, namely 
symmetry with respect to the line 
joining both star. It is thus important, to get additional complementary 
information about the non-illuminated part of the secondary star and to 
combine both aspects.

\begin{figure}[t]
\plotfiddle{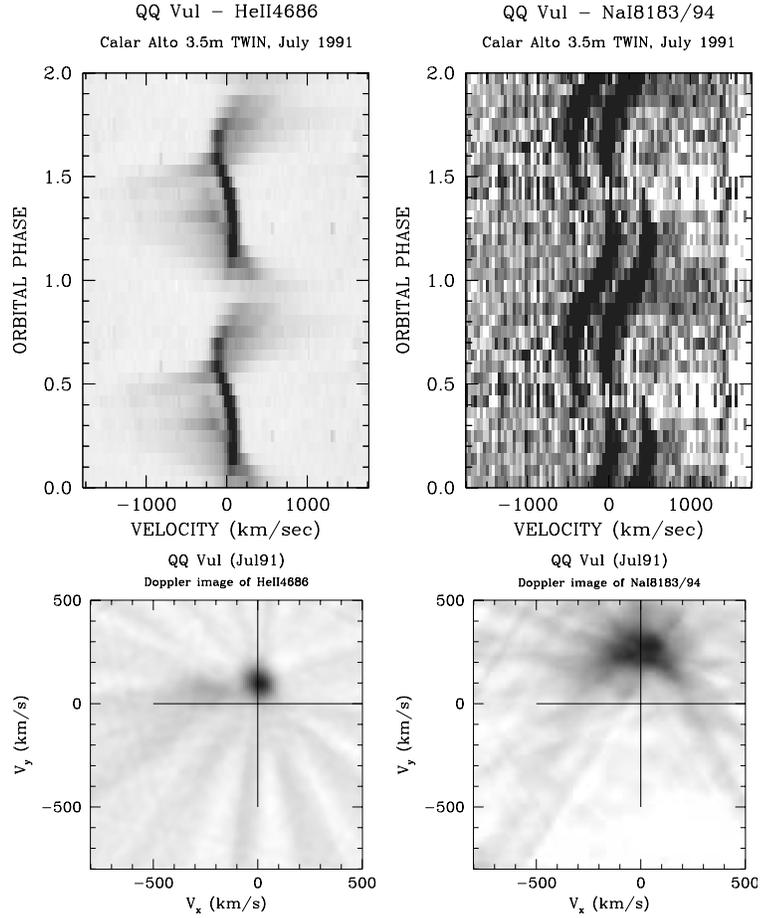}{11cm}{0}{75}{75}{-200}{-170}
\caption{Trailed spectrograms of the HeII 4686 emission and the NaI 8183/94
absorption lines of QQ Vul obtained simultaneously in a 1991 high state.
The spetra were continuum-subtracted for representation in the figure.
The Doppler maps below show emission from the illuminated front-side
and absorption mainly from the non-illuminated back side of the secondary star.
}
\label{qq_trails}
\end{figure}

In Fig.~\ref{qq_trails} an example is shown where quasi-chromospheric 
reprocessed emission lines and photospheric absorption lines has been 
recorded simultaneously in one system, QQ Vulpeculae. The blue trailed 
spectrogram shows essentially the same components as seen above for the 
case of HU Aqr. Here we emphasize the occurence of the bright NEL 
originating from the secondary star. Line emission from the stream can not
be traced very well in the Doppler map due to the rather coarse phase 
resolution of this data set ($\Delta \phi \simeq 0.05$). The NEL is visible 
for about 60\% of the orbital cycle and has a radial velocity amplitude
of $\sim$120 km s$^{-1}$. 

The photospheric Na{\sc i}-doublet at 8183/8194\AA\ is visible at complementary
phases to that of He{\sc ii} 4686 and possesses a higher radial velocity 
amplitude of $\sim$265 km s$^{-1}$. The Doppler images of both lines 
show a distinct separation between  emission and absorption features. Emission 
was of course not expected from the non-illuminated backside of the secondary
star but it was not clear at all that absorption is suppressed so radically 
from the frontside. A similar effect, perhaps not as pronounced as in
QQ Vul, has been seen in AM Her by Southwell et al.~(1995) and used to map 
the Na{\sc i} surface distribution by Davey \& Smith (1996). 
These two examples demonstrate that care must be taken when using observed 
radial velocity amplitudes in terms of {\it bona fide} uniformly 
distributed photospheric absorption lines in order to 
estimate stellar masses of polars. The effect of illumination 
on the photospheric structure and its relation to mass transfer (and mass
transfer variations on short timescales) is an unexplored field.

\section{Summary and outlook}
Using some selected systems we have demonstrated how multi-epoch, 
multi-wavelength observations have allowed new and detailed insights
into the accretion phenomena in AM Herculis stars. In the foreseeable 
future mapping of even smaller substructures will become possible. 
XMM will probably allow to measure the sizes of hard and soft X-ray emissions
separately, 8m-class optical telescopes will help to explore the 
accretion stream in the magnetosphere in even greater detail by 
constructing eclipse maps in velocity bins.

\acknowledgments
This work was supported by the BMB+F under DARA grants 50 OR 9403 5 and
50 OR 9706 8.


\begin{references}
\reference 
	Bailey J., Cropper M., 1991, \mnras\ 253, 27
\reference 
	Beuermann K., 1997, {\em Perspectives in High Energy Astronomy \& 
	Astrophysics}, Tata Institute of Fundamental Research, Mumbai, India, 
	August 12-17, 1996
\reference 
	Beuermann K., Burwitz V., 1995, ASP Conf.~Ser.~85, 99
\reference 
	Beuermann K., Schwope A.D., 1989, \aap\ 223, 179
\reference 
	Ciardi D., Howell S.B., Saxton, 1997, this volume
\reference 
	Cropper M., 1987, \mnras\ 228, 389
\reference 
	Davey S.C., Smith R.C., 1996, \mnras\ 280, 481
\reference 
	Diaz M.P., Steiner J.E., 1994, \aap\ 283, 508
\reference
	Ferrario L., BAiley J., Wickramasinghe D.T., 1996, \mnras\ 282, 218
\reference
	Hakala P.J., 1995, \aap\ 296, 164
\reference
        Hakala P.J., Watson M.G., Vilhu O., Hassall B.J.M., Kellett B.J.,
        Mason K.O., Piirola V., 1993, \mnras\ 263, 61 
\reference
	Hameury J.-M., King A.R., 1988, \mnras\ 235, 433
\reference 
	Marsh T.R., Horne K., 1988, \mnras\ 235, 269
\reference 
	Piirola V., Reiz A., Coyne G.V., 1987, \aap\ 185, 189
\reference 
	Schwope A.D., 1996, in {\em Cataclysmic Variables and Related 
	Objects}, A.~Evans and J.H.~Wood (eds.), Kluwer, Dordrecht, p.~189
\reference 
	Schwope A.D., Beuermann K., 1989, \aap\ 222, 132 
\reference 
	Schwope A.D., Beuermann K., Jordan S., 1995, \aap\ 301, 447
\reference 
	Schwope A.D., Mantel K.-H., Horne K., 1997, \aap\ 319, 894
\reference
	Schwope A.D., Thomas H.-C., Beuermann K., 1993, \aap\ 271, L25 
\reference
	Schwope A.D., Beuermann K., Burwitz V., Mantel K.-H., Schwarz R., 
	1995a, {\em Proc.~Padua Conference 
        on CVs and related Physics}, eds.~A.~Bianchini et al., Kluwer, p.~389
\reference
	Schwope A.D., Thomas H.-C., Beuermann K., Burwitz V., Jordan S., 
        Haefner R., 1995b, \aap\ 293, 764
\reference 
	Shafter A.W., Reinsch K., Beuermann K., Misselt K.A., Buckley D.A.H.,
        Burwitz V., Schwope A.D., 1995, \apj\ 443, 319
\reference 
	Sirk M., Howell S.B., 1997, this volume
\reference 
	Sohl K.B., Watson M.G., Rosen S.R., 1995, ASP Conf.~Ser.~85, 306
\reference 
	Southwell K., Still M.D., Smith R.C., Martin J.S., 1995, \aap\ 302, 90
\reference 
	Stockman H.S., Schmidt G.D., 1996, \apj\ 468, 833
\reference 
	Warren J.K., Sirk M.M., Vallerga J.V., \apj\ 445, 909
\reference 
	Watson M.G., 1993, Adv.~Space Res.~13(12), p.~125
\end{references}
\end{document}